# A Study of Improved Limiter Formulations for Second-Order Finite Volume Schemes Applied to Unstructured Grids

**Frederico Bolsoni Oliveira**

Instituto Tecnológico de Aeronáutica – DCTA/ITA, 12228-900, São José dos Campos, SP, Brazil.
fredericobolsoni@gmail.com

**João Luiz F. Azevedo**

Instituto de Aeronáutica e Espaço – DCTA/IAE/ACE-L, 12228-904, São José dos Campos, SP, Brazil.
joaoluiz.azevedo@gmail.com

***Abstract.*** *A general, compact way of achieving second-order in finite-volume numerical methods is to perform a MUSCL-like, piecewise linear reconstruction of flow properties at each cell interface. To avoid the surge of spurious oscillations in the discrete solution, a limiter function is commonly employed. This strategy, however, can add a series of drawbacks to the overall numerical scheme. The present paper investigates this behavior by considering three different limiter formulations in the context of a second-order, finite volume scheme for the simulation of steady, turbulent flows on unstructured meshes. Three limiter formulations are considered: the original Venkatakrishnan limiter, Wang's modification to the Venkatakrishnan limiter and Nishikawa's recently introduced $R_3$ limiter. Three different configurations of the fully-developed, two-dimensional, transonic NACA 0012 airfoil are analyzed, configured with different angles of attack and similar freestream properties. The gas dynamics are modeled using the Reynolds-averaged Navier-Stokes (RANS) equations, where the negative Spalart-Allmaras turbulence model is used to solve the closure problem. All limiters are shown to yield similar results for all configurations of this case, although with different dissipative characteristics, provided their control constants are used within appropriate intervals. The presented numerical results are in good agreement with experimental data available in the literature.*

## 1. INTRODUCTION

The usage of second-order schemes for discretely solving the gas dynamics equations is currently considered standard practice in the industry. In the context of the finite volume numerical method, a general, compact way of achieving second-order is to perform a piece-wise linear, MUSCL-like, reconstruction of flow properties at each cell interface (van Leer, 1979; Barth and Jespersen, 1989). This type of reconstruction, however, is known to induce spurious oscillations near solution discontinuities, such as shock waves. Hence, simulations of transonic and supersonic flows are particularly affected by this behavior. A common strategy to solve this problem is to make use of a limiter function (van Leer, 1977). This strategy, however, can add a series of drawbacks to the overall numerical scheme, mainly the introduction of additional artificial dissipation in undesired regions of the domain and numerical difficulties in the convergence of the residue (Venkatakrishnan, 1993, 1995).

For general aeronautical applications, it is of interest to the user that the numerical schemes employed in the simulation are capable of being utilized with unstructured meshes. Thus, more options will be available to the CFD user. The first well-known effort to develop a limiter formulation that is applicable to general unstructured meshes is due to Barth and Jespersen (1989). This formulation was later improved by Venkatakrishnan (1995), which also included an additional term to avoid numerical complications in regions where the solution is near constant. Further efforts were made in order to expand the limiter formulation to high-order schemes, particularly by Michalak and Ollivier-Gooch (2008), but under a completely different framework in relation to the previous approaches. Recently, Nishikawa (2022) presented new limiter functions that are suitable for usage with schemes of up to fifth-order. However, different from Michalak and Ollivier-Gooch (2008), Nishikawa's formulation preserves the same framework employed by Venkatakrishnan (1995). Therefore, it is theoretically easier to port the new formulation to existing CFD codes.

Since Nishikawa's work was published relatively recently, the authors are unaware of any further investigations regarding the behavior of the new family of limiters when used to solve complex problems. Moreover, the original publication dealt with high-order schemes, so a natural extension would be to assess the quality of the numerical solutions that are obtained when the new limiters are used in conjunction with low-order formulations. The present work seeks to fulfill this need by evaluating the performance of Nishikawa's $R_3$ limiter against three different configurations of a fully developed, transonic, turbulent case of interest to the aeronautical field. The case is the transonic NACA 0012 airfoil (McDevitt and Okuno, 1985) with three different angles of attack. Other freestream properties are kept approximately constant among all configurations. The $R_3$ limiter is implemented in the *Instituto de Aeronáutica e Espaço* (IAE) CFD code, BRU3D (Bigarella and Azevedo, 2009, 2012; Oliveira and Azevedo, 2022, 2023), which employs a cell-centered, second-order implicit formulation for the discretization of all conservation laws. Inviscid numerical fluxes of the Reynolds-Averaged

Navier-Stokes equations (RANS) are reconstructed using Roe's flux difference splitting scheme (Roe, 1981). Moreover, the closure problem, inherent to the RANS formulation, is solved by using the negative Spalart-Allmaras turbulence model, SA-neg (S. R. Allmaras and Spalart, 2012). A second-order, total-variation diminishing (TVD) formulation is achieved by performing a piecewise linear reconstruction of the solution inside each computational cell, coupled with the usage of a limiter. In order to better address the capabilities of the $R_3$ limiter, two other limiters are also considered in this study. The first one is the original Venkatakrishnan limiter (Venkatakrishnan, 1995). The second one is Wang's modification to the Venkatakrishnan limiter (Wang, 2000). Throughout the paper, comparisons are made between the results from the three limiters and experimental data. Further sections of this document define the numerical formulations employed in the current effort, followed by the presentation and discussion of the obtained results. Finally, the authors provide concluding remarks.

## 2. NUMERICAL FORMULATION

The present paper makes use of an unstructured, cell-centered, finite volume (FV) discretization scheme to numerically solve the system of conservation laws of interest. A more in-depth description regarding the overall numerical formulation used throughout the work is given by the authors in previous publications (Oliveira and Azevedo, 2022, 2023) and, thus, we refer the interested reader to those references. As such, the current section will focus entirely in presenting the numerical formulation of each limiter.

In the FV framework, the domain is subdivided into multiple small control volumes, called cells, where the conservation laws must hold. Considering that the $i$-th cell of the computational domain has $n_f$ faces and a volume equal to $\mathbb{V}_i$, one can write the discretized integral form of the RANS equations applied to $\mathbb{V}_i$ as

$$\frac{\partial \vec{Q}_i}{\partial t} = -\frac{1}{\mathbb{V}_i} \sum_{k=1}^{n_f} \left( \vec{\mathcal{F}}_k \cdot \vec{S}_k \right). \tag{1}$$

In Eq. (1), $\vec{Q}$ is the vector of conserved properties, $t$ is the time variable and $\vec{\mathcal{F}}$ is a vector of geometric flux vectors. Furthermore, $\vec{S}_k$ is the area vector of the $k$-th face of the current $i$-th cell, pointing outwards. During the development of the scheme, since the reconstructed solution is taken to be discontinuous across different control volumes, it becomes necessary to define a common numerical flux at the interfaces of each cell, $\vec{\mathcal{F}}_k$. Here, Roe's numerical flux is utilized (Bigarella and Azevedo, 2009; Roe, 1981). The evaluation of a common interface flux is generally a function of reconstructed properties on both sides of the interface. By adopting a MUSCL-like (van Leer, 1979) strategy, a second-order scheme is obtained by making use of a piecewise linear reconstruction, as defined by Barth and Jespersen (1989). If the $j$-th cell is the direct neighbor of the $i$-th cell at its $k$-th face, then a generic property, $b$, can be reconstructed as follows

$$b_{ki} = b_i + \psi_i \left( \overrightarrow{\nabla b}_i \cdot \vec{r}_{ki} \right), \tag{2}$$

and

$$b_{kj} = b_j + \psi_j \left( \overrightarrow{\nabla b}_j \cdot \vec{r}_{kj} \right), \tag{3}$$

where $b_{ki}$ is the reconstructed property vector associated with the $i$-th cell and evaluated at the $k$-th face centroid. Likewise, $b_{kj}$ is the reconstructed property vector associated with the $j$-th cell and evaluated at the same face centroid. Variables $\vec{r}_{ki}$ and $\vec{r}_{kj}$ are distance vectors that point from the $i$-th and $j$-th cell centroid to the current face centroid, respectively. Lastly, $\psi_i$ and $\psi_j$ are the $b$ property limiters associated with each cell.

Three different limiter formulations, applied to unstructured meshes, are considered in the present effort. The first one is Venkatakrishnan's limiter (Venkatakrishnan, 1995), $\psi^V$. In order to compute it, it is necessary to evaluate property variations at each one of the current cell faces, as shown below

$$\psi_k^V = \frac{\left( \Delta_+^2 + \epsilon^2 \right) + 2\Delta_-\Delta_+}{\Delta_+^2 + 2\Delta_-^2 + \Delta_+\Delta_- + \epsilon^2}, \tag{4}$$

where $\Delta_+$ and $\Delta_-$ are property variations defined as

$$\Delta_- \equiv \overline{b_{ki}} - b_i \tag{5}$$

and

$$\Delta_+ \equiv \begin{cases} b_{\max} - b_i, & \text{if } \overline{b_{ki}} \geq b_i, \\ b_{\min} - b_i, & \text{if } \overline{b_{ki}} < b_i, \end{cases} \quad (6)$$

with

$$\overline{b_{ki}} = b_i + \overrightarrow{\nabla b_i} \cdot \vec{r}_{ki}. \quad (7)$$

In Eq. (4), $b_{\max}$ and $b_{\min}$ are the maximum and minimum discrete $b$ property values among the current cell and all of its directly adjacent neighbors. Also, notice that Eq. (7) is nothing more than an unlimited linear property reconstruction. The parameter $\epsilon^2$ is a modification, introduced by Venkatakrishnan, to avoid numerical complications in nearly constant regions. It is defined as

$$\epsilon^2 = (\epsilon_V \Delta x_i)^3, \quad (8)$$

in which $\epsilon_V$ is a constant, assuming values in the range $\epsilon_V \in [0.01, 10]$, and $\Delta x$ is a characteristic length scale of the current cell. The definition of the latter is completely open to interpretation, but it is commonly taken to be $\Delta x \equiv \sqrt[3]{\mathbb{V}_i}$, which is the one employed here. The cell limiter is, then, computed simply as

$$\psi_i^V = \min\left(\psi_k^V\right). \quad (9)$$

The original definition of the $\epsilon^2$ term from Venkatakrishnan's formulation can introduce computational problems due to the drastic differences in value that can exist when the biggest and smallest cell sizes of the domain are taken into account. In an attempt to address this issue, Wang (2000) proposed the following definition for $\epsilon^2$:

$$\epsilon^2 = \left[\epsilon_W \left(b_{gl.\,\max} - b_{gl.\,\min}\right)\right]^2. \quad (10)$$

In Eq. (10), $\epsilon_W$ is a constant whose recommended values lie in the range $\epsilon_W \in [0.01, 0.20]$. Furthermore, $b_{gl.\,\max}$ and $b_{gl.\,\min}$ are the global maximum and minimum properties over the entire computational domain. In order to differentiate the resulting modified limiter from the previous one, the authors will refer to it as the $\psi^W$ limiter.

The last limiter is part of the $R_p$ family of limiters introduced by Nishikawa (2022), particularly the one obtained when $p$ is made equal to 3. The Nishikawa $R_3$ limiter is calculated as follows:

$$\psi_k^{R_3} = \begin{cases} \frac{(|\Delta_+|^3 + \epsilon^3) + |\Delta_+| S_3}{(|\Delta_+|^3 + \epsilon^3) + |\Delta_-|(\Delta_+^2 + S_3)}, & \text{if } |\Delta_+| \leq 2|\Delta_-|, \\ 1, & \text{if } |\Delta_+| > 2|\Delta_-|, \end{cases} \quad (11)$$

with

$$S_3 = 4|\Delta_-|^2. \quad (12)$$

Once again, a new definition for the $\epsilon$ term is utilized, as specified below:

$$\epsilon^3 = (\epsilon_{R_3} \Delta x)^4. \quad (13)$$

Optimal values for the $\epsilon_{R_3}$ parameter are not fully defined. However, since the introduction of this term follows a similar rationale to Venkatakrishnan's implementation, it is expected that the utilization of values in the same range of $\epsilon_V$ might give good results. Limiter cell values are computed in an equivalent manner to Eq. (9).

In theory, $\psi^{R_3}$ can preserve up to third-order of spatial accuracy in smooth regions of the flow for numerical schemes that can attain such property (Nishikawa, 2022). In the present case, the overall numerical scheme is only up to second-order accurate. Thus, the usage of $\psi^{R_3}$ might not introduce sufficient amounts of artificial dissipation to sustain stable behavior in specific applications. This, however, is not the case for the simulations performed here.

The steady state solution is obtained by integrating the time derivative using an implicit Euler scheme. The resulting linear system is solved by using the restarted generalized minimum residue (GMRES($m$)) iterative method, where the solution is restarted after every $m$ Arnoldi iterations. Here, $m$ is taken to be equal to 200. This discretization process often results in ill-conditioned linear systems. Consequently, the usage of a preconditioner is required for achieving convergence in the solution of the linear system. Since the current implementation is parallelized using domain decomposition and operates on distributed memory, the Additive Schwartz preconditioner (Cai and Sarkis, 1999) is applied to the entire linear system, together with an incomplete lower-upper factorization with fill level 3 (ILU(3)) that acts on each subdomain separately (Witherden *et al.*, 2017; Hicken *et al.*, 2011). In order to quickly achieve steady state, a variable local time step approach is employed by keeping the Courant-Friedrichs-Lewy (CFL) number constant throughout the entire solution domain per nonlinear iteration. For all simulations, Harten's entropy fix is set to $0.125$ and the viscous fluxes are evaluated using the $LJ0$ scheme (Oliveira and Azevedo, 2023).

## 3. DESCRIPTION OF TEST CASES

The solutions obtained by employing the three different limiter formulations are assessed in the context of three different configurations of the transonic NACA 0012 airfoil case. Each configuration is subject to similar freestream Reynolds number, $Re$, and Mach number, $M$, but different angles of attack, $\alpha$. The case setup follows the available experimental data from McDevitt and Okuno (1985) and, therefore, consistency among each configuration freestream property is subject to experimental variation. Table 1 shows the freestream state of each configuration, where $T_\infty$ is the temperature of the flow and $c$ is the reference chord length. Small differences in Reynolds and Mach numbers among the three configurations stem from the absence of full control over the experimental variables.

Table 1. Freestream conditions for the three transonic NACA 0012 test cases.

| Config. | $Re_\infty$ | $M_\infty$ | $T_\infty$ | $c$ | $\alpha$ |
|---|---|---|---|---|---|
| 1 | $6.5 \cdot 10^6$ | 0.803 | 300 K | 1 m | $-0.10$ deg. |
| 2 | $6.1 \cdot 10^6$ | 0.804 | 300 K | 1 m | $+0.96$ deg. |
| 3 | $6.0 \cdot 10^6$ | 0.778 | 300 K | 1 m | $+2.03$ deg. |

Since BRU3D implements a three-dimensional formulation, two-dimensional behavior is simulated using a three-dimensional domain, with a single cell in the depth, or spanwise, direction. A constant mesh is used throughout all configurations, which is available at NASA's Turbulence Modeling Resources website, TMR (Rumsey, 2022), under the family 1 sequence of meshes. It is a C-shaped mesh composed of 1792 by 512 hexahedra, which simulates quadrilateral cells in the present case. Cells are clustered near the wall in order to keep $y^+ < 1$ at all locations directly above the airfoil surface. Figure 1 provides an overview of the NACA 0012 mesh, including a zoomed-in view of the region that surrounds the airfoil surface.

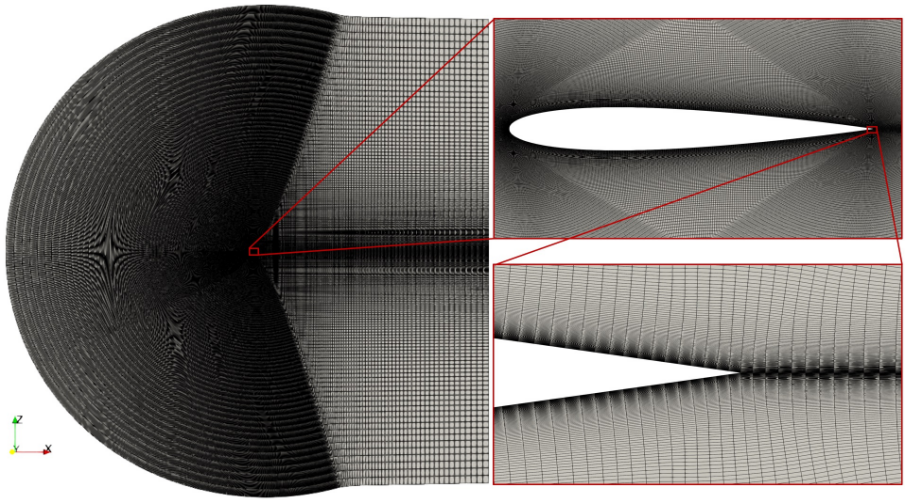

Figure 1. Overview of the mesh used in the transonic NACA 0012 cases. Focus is given to the region near the airfoil surface. Mesh obtained from TMR (Rumsey, 2022).

Boundary conditions for this case follow the standard setup for airfoil simulation. That is, the airfoil surface is set as an adiabatic wall, while the outer surface is configured as a non-reflective farfield based on the propagation of Riemann invariants. Since this is a transonic case, shock waves are expected to develop in the domain and interact with the boundary layer, providing a nice challenge to the limiters.

## 4. RESULTS AND DISCUSSION

### 4.1 Convergence Criteria

Unless otherwise stated, results are given in a converged state throughout this section. Steady state is calculated by dynamically ramping up the global CFL number from 0.1 to 10000. In the current study, a case is considered converged if a reduction of at least 6 orders of magnitude is observed in the residue associated with the continuity equation. This definition is significantly "relaxed" in relation to research previously done by the authors (Oliveira and Azevedo, 2022, 2023), in which a decrease of at least 10 orders of magnitude was sought. The presence of strong shock waves in the numerical solution, together with the usage of a very fine mesh, have significantly decreased the convergence rate of the numerical schemes in use. Hence, a less strict convergence criteria is adopted. All limiters herein considered are capable of achieving this behavior when configurations 1 and 2 are computed. However, all of them suffer from convergence stall when configuration 3 is simulated. Figure 2 depicts this behavior by plotting the continuity equation residue, normalized by the first iteration residue value, for different combinations of case configuration and limiter formulation. The convergence problem, in the current implementation, appears to be due to the usage of a limited piecewise linear reconstruction together with a reasonably high CFL value. The residue related to the turbulence model, not pictured here, is particularly affected by these numerical difficulties due to its slow and chaotic convergence pattern. If the limiter is taken to be zero everywhere, the scheme decays to first-order and convergence to machine zero becomes attainable, albeit very slowly, as shown in Fig. 2 for the first 5000 iterations, and in Fig. 3 for the complete convergence to machine zero. Convergence to machine zero appears to also be possible for the limited second-order scheme, provided a very small CFL value is used, similar in magnitude to what one would use in conjunction with an explicit scheme. This behavior could be due to the simplifications introduced in the calculation of the approximate Jacobian matrix used in the present implicit time march, which might be unable to capture the full nonlinear characteristics of the limited numerical scheme. The investigation of this hypothesis, however, is beyond the scope of the present study. An unlimited scheme ($\psi = 1$), on the other hand, becomes completely unstable when solving the configurations analyzed here. Under such condition, very strong spurious oscillations develop around the shock wave, as expected, which quickly make the scheme diverge. Figure 4 shows a temperature contour for a situation in which the spurious oscillations are starting to develop, and the solution has not yet diverged, making them particularly visible in the region aft of the shock, as the oscillations are transported by the fluid flow and the affected region grows.

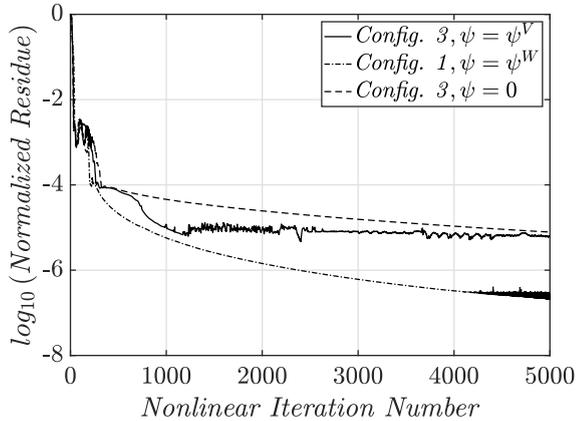

Figure 2. Normalized continuity equation residue obtained with different limiters and case configuration.

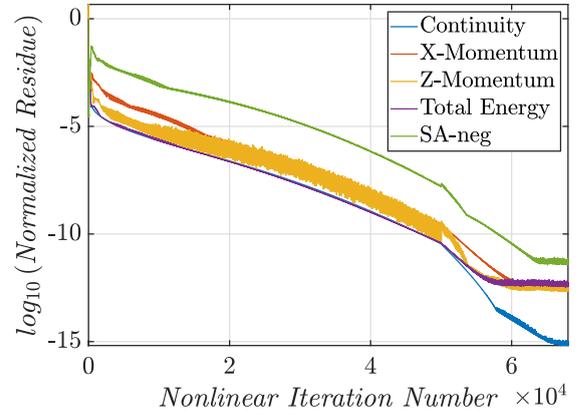

Figure 3. Normalized residue obtained with the limiter set to zero in the entire domain (first order scheme) when solving configuration 3.

### 4.2 Sensitivity to Control Parameters

Another important factor to consider when performing this analysis is the solution sensitivity to the control parameter values of each limiter formulation. The three limiters, $\psi^V$, $\psi^W$ and $\psi^{R_3}$, are controlled by the constants $\epsilon_V$, $\epsilon_W$ and $\epsilon_{R_3}$, respectively. As described previously, the main goal behind the introduction of these parameters is to avoid numerical problems in the limiter calculation at regions of near constant property values. Besides avoiding the obvious division by zero, the extra terms also theoretically allow the solution to converge to a machine zero steady state by constraining the region in which the limiter assumes a value different from 1. Rigorously, the resulting numerical formulation is only TVD if $\epsilon$ is taken to be zero. For all three formulations, $\epsilon$ values that are higher than zero weakens the TVD definition, allowing for some oscillations to occur near numerical discontinuities by reducing the region in which the limiter effectively becomes "active". This, in turn, reduces the non-linearity of the numerical scheme, which theoretically allows for a more seamless convergence of the solution.

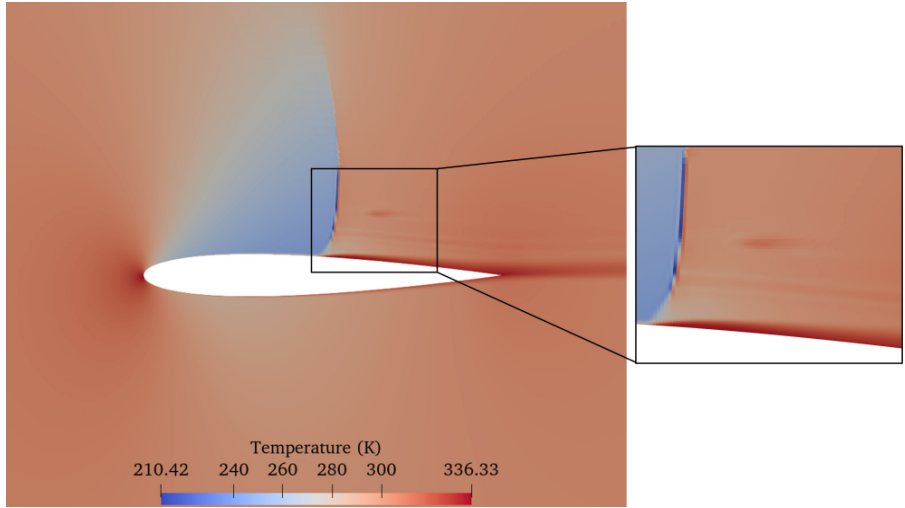

Figure 4. Temperature contours for configuration 3, run in second-order without a limiter for 500 iterations. Spurious oscillations develop in the domain, making this particular numerical scheme unstable.

In this subsection, the authors assess the dependency of the numerical solution on the value of the control parameters by measuring the changes in the predicted macro properties of the flow, which in the current case are the airfoil aerodynamic coefficients, when the three limiters are used with different values of $\epsilon_V$, $\epsilon_W$ and $\epsilon_{R_3}$. Figure 5 shows the lift and drag coefficients, $C_L$ and $C_D$, computed for configuration 3 when using multiple control parameter values within their recommended range. Wang's limiter is the only one that has been evaluated with an $\epsilon$ value exactly equal to zero. However, due to the nature of that limiter formulation, $\psi^W = \psi^V$ when $\epsilon = 0$.

Since, for this configuration, no formal convergence is attained, the numbers displayed result from an average performed over multiple iterations. First, the residue is decreased as much as possible, allowing the aerodynamic forces to exhibit a high-frequency, low amplitude, seemingly random, non-physical oscillatory behavior, typical of a solution that has not properly converged to machine zero. At this point, an average is taken of the aerodynamic coefficients along 5000 iterations, in which the aerodynamic coefficients are computed after each 500 iterations. The authors emphasize that the amplitude of the oscillations are, indeed, very low. For instance, on the worst observed case, the amplitude of the oscillations of the drag coefficient are of the order of 0.0001 (i.e., 1 drag count).

As one can see from Fig. 5, the aerodynamic coefficients calculated for the most critical setup of the current case (configuration 3) experience negligible change to modifications made to the limiter control parameter values, provided the $\epsilon$'s are constrained to their recommended intervals. Among all calculated values, no change higher than 0.40% is observed for the computed lift coefficient. Likewise, no difference higher than 0.27% is observed for the drag coefficient. Of course, simple integral property values, such as the aerodynamic coefficients, do not paint the complete picture regarding the differences in the solution field calculated using the three limiters. As such, the proceeding subsection of this paper is dedicated to analyze the differences observed in the flow property values when each limiter is used. For those analyses, the limiter control parameter values used are shown in Tab 2.

Table 2. Limiter control parameter values used in the reported results.

| Control Parameter | Value |
|---|---|
| $\epsilon_V$ | 5.0 |
| $\epsilon_W$ | 0.1 |
| $\epsilon_{R_3}$ | 5.0 |

Another important pattern that can be observed in Fig. 5 is the overall tendency of the aerodynamic forces to increase, even though very slightly, when $\epsilon$ is increased from $\epsilon = 0$. This has been the case for every limiter analyzed, except for the drag coefficient computed with $\psi^W$, which decreases when $\epsilon_W$ is increased near zero. However, the change in drag coefficient between $\epsilon_W = 0$ and $\epsilon_W = 0.02$ is of only 0.000003, which is within the uncertainty generated by the unconverged numerical solution. In other words, a difference of only 0.03 of a drag count is certainly within the realm of the possible change in the drag coefficient if the numerical solutions were forced to machine zero in the present calculations. The authors felt that the computational costs of such exercise were not justifiable in the present study.

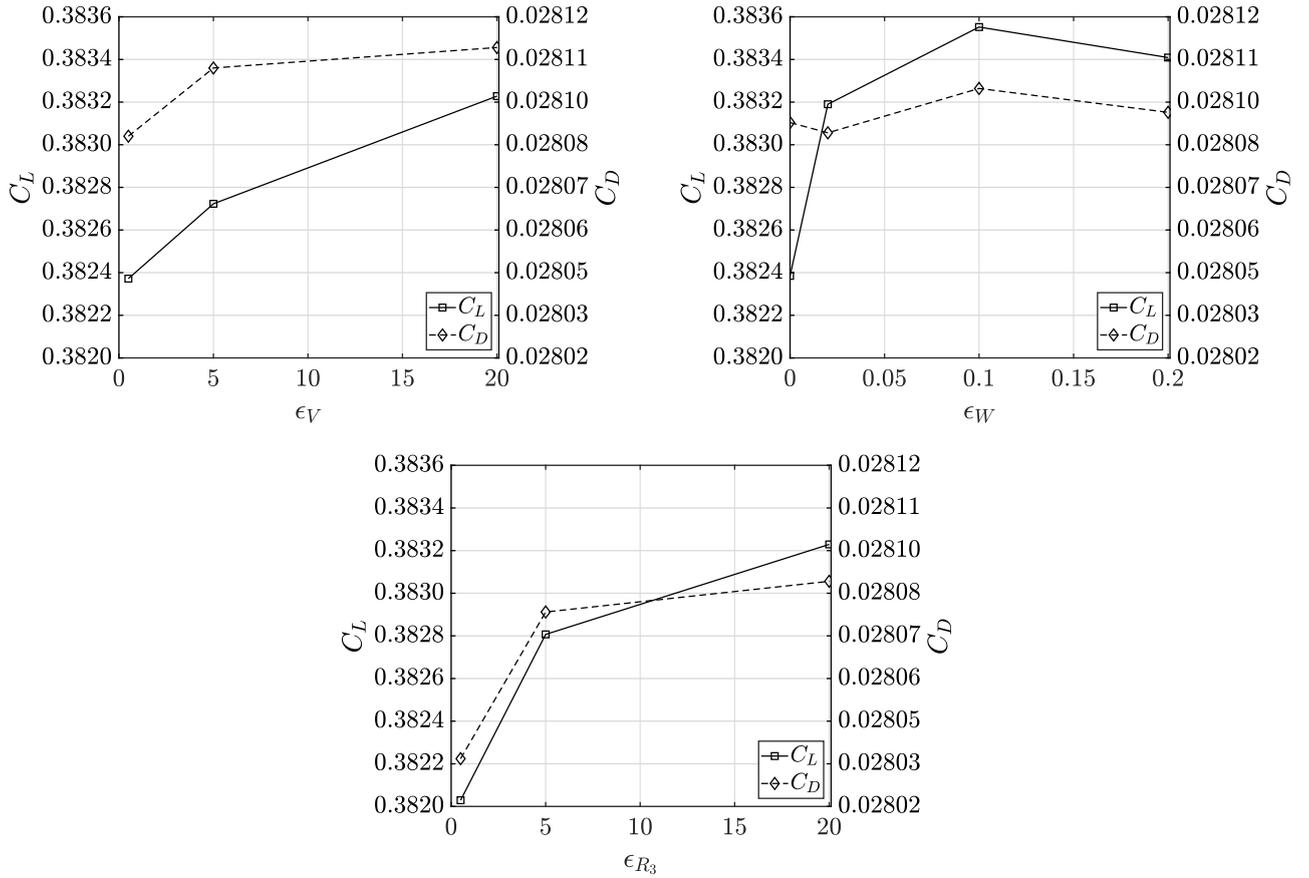

Figure 5. Sensitivity of the aerodynamic coefficients, computed using configuration 3, to the control parameters $\epsilon_V$, $\epsilon_W$ and $\epsilon_{R_3}$ of the $\psi^V$, $\psi^W$ and $\psi^{R_3}$ limiters, respectively.

### 4.3 Pressure Field and Pressure Limiter Data

In the present subsection, the authors aim to provide the reader with a thorough understanding of the effects that the limiter formulation has over the solution field. Experimental data is only available for the pressure field, as published by McDevitt and Okuno (1985). Therefore, the current analysis is constrained to that field variable only.

Figure 6 displays the pressure contours of each test case, calculated using Venkatakrishnan's limiter, in order to illustrate the overall behavior of the solution. Since the pressure limiter acts directly over the pressure field, this particular figure is scaled according to the pressure value, and not by its dimensionless counterpart, the pressure coefficient, $C_p$, which is commonly used in the literature for this type of plot. Results obtained using the other two limiters are virtually identical at this plot scale and, therefore, are not shown. The aerodynamic forces that act on the airfoil are partially dependent on the pressure distribution along its surface. Hence, a better insight can be gained regarding the results shown in the previous subsection by plotting, in Fig. 7, the estimated pressure coefficient along the airfoil surface for all case configurations. In order to allow further evaluation of the obtained results, experimental data from McDevitt and Okuno (1985) are also shown. From that figure, it becomes clear that, for this particular case, all three limiters lead to the same converged result, no matter the severity and location of the shock wave. No notable differences are observed even in the region that directly surrounds the shock waves, as seen in the zoomed-in views available in the same figure. The same observation is also valid for all the other flow primitive properties, which explains the very small variance in the predicted lift and drag coefficients presented in the previous subsection for test case 3, which, as seen here, is a conclusion that can be extended to all the other test cases.

With respect to the quality of the computed results in comparison to experimental data, it is evident from Fig. 7 that the overall physics of the flow are correctly captured by the numerical simulation, especially for test cases 1 and 3. For those test cases, the numerical schemes can accurately determine the average pressure distribution throughout the entire extension of the airfoil, including the shock location and behavior. The same observation is also true, to a certain extent, for test case 2. For this test case, the computed $C_p$ shows very good agreement with experimental data for the region that goes from the airfoil leading edge to the onset of the shock wave, on both the upper and lower surfaces of the airfoil. The same can also be said for the last 20% of the airfoil chord. The complex behavior that develops between these two regions, however, is not well represented by the simulation. The property jump in the upper surface shock appears to be overpredicted by the numerical simulation, although its location is correctly captured. Furthermore, the numerical data point towards the existence of a well-defined shock wave at the lower surface of the airfoil, while the experimental measurements display a smoothed-out shock, diffused across approximately 20% of the airfoil chord, from $x/c = 0.4$ to $x/c = 0.6$. These discrepancies are likely due to the turbulence model used, the SA-neg, which appears to be unable to correctly model the complex shock wave/boundary-layer interaction that develops in the airfoil lower surface. The discretization itself does not appear to be a source of error in this instance, as the solution is currently mesh independent and insensitive to changes made to the scheme control parameters, such as the limiter formulation and entropy fix definition, with the latter not being a subject of the present effort. This conclusion is, of course, only valid considering that no other source of disturbance exists in the experimental data.

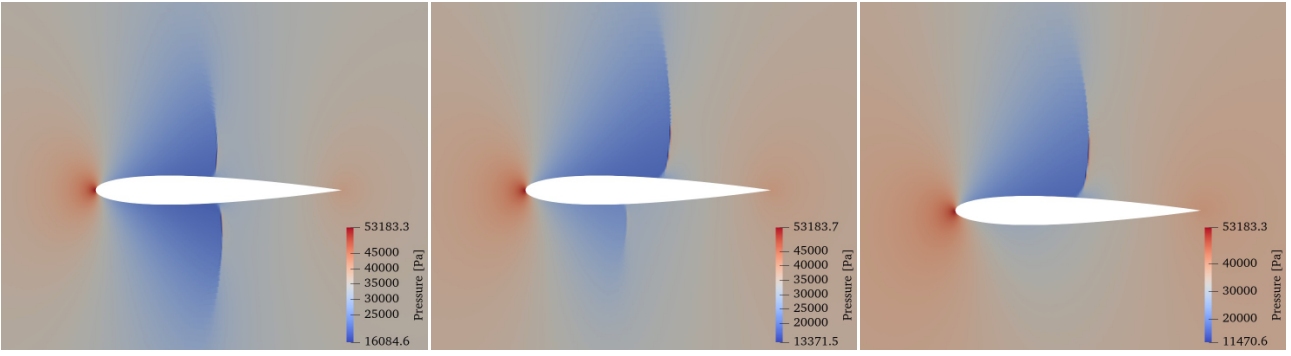

Figure 6. Pressure contours for test cases 1, 2 and 3, respectively, computed using Venkatakrishnan's limiter. Results for the other limiters are virtually identical at this plot scale.

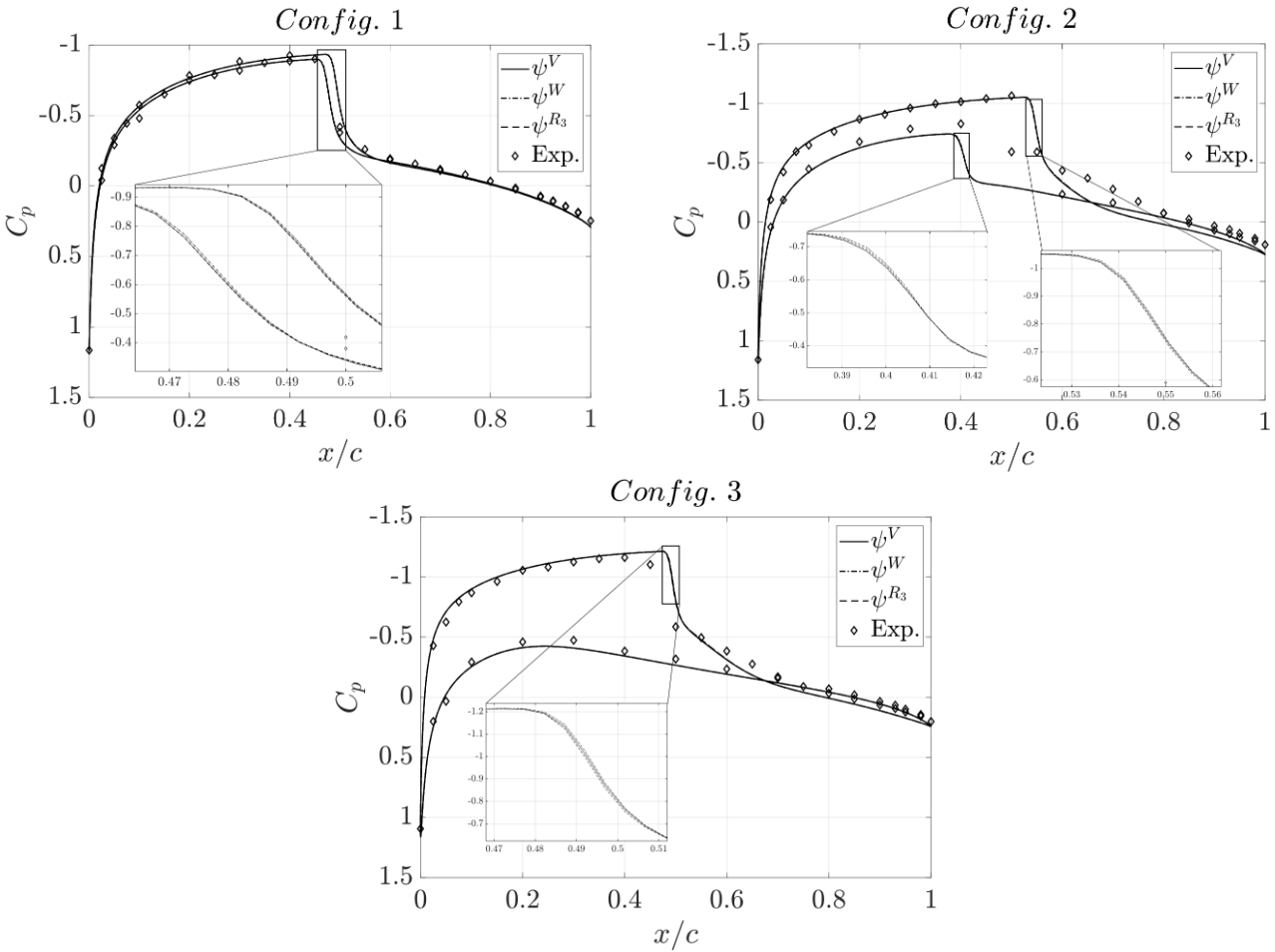

Figure 7. Surface pressure coefficient, $C_p$, for test cases 1, 2 and 3, respectively. Experimental data from McDevitt and Okuno (1985) are included for comparison.

Despite the fact that the results associated with the physical properties of the flow are virtually identical across all formulations, there are discernible differences in the computed limiter values, which can be assessed with the help of a contour plot. Figure 8 shows the pressure limiter contours for all possible combinations of test cases and limiter formulations, totaling 9 different subplots. Results from each row are obtained using the $\psi^V$, $\psi^W$ and $\psi^{R_3}$ limiters, respectively. Likewise, results from each column refer to test cases 1, 2 and 3, respectively. The first notable characteristic of the presented solutions is that, under no circumstance, the limiter value becomes greater than 1 anywhere in the domain. By construction, Venkatakrishnan's limiter and, by extension, Wang's modified formulation, can assume values higher than one in order to keep the second-order scheme within its monotonicity bounds while also maintaining desirable numerical features, such as continuous differentiability. On the other hand, Nishikawa's limiter, by construction, is able to keep these desirable features while never exceeding 1. Nevertheless, all limiters are observed to be constrained to values below 1 in the present test cases. Both $\psi^V$ and $\psi^{R_3}$, computed using the control parameter values shown in Tab 2, have a "region of influence" which includes the near wall region. That is, these limiters assume non-unitary values in the region where the shock wave touches the airfoil surface. This is not the case for the $\psi^W$ limiter, which is significantly less restrictive in the near wall region. Another important feature that is also apparent in Fig. 8 is the more dissipative nature of the Venkatakrishnan limiter, illustrated by the fact that its non-unitary region is visibly larger than the other two for all three test cases. The less dissipative nature of the $R_3$ limiter, which is expected due to the fact that it is suitable to be used in conjunction with high-order schemes, can also be seen in the same figure. Values close to zero are only achieved in the two cells that share a common face, width-wise, past the shock wave, a feature displayed by all limiters in the current context. However, only the $R_3$ limiter quickly returns to its unlimited state in the other neighboring cells, with the other two limiters still presenting reasonably low limiter values for at least one extra cell in each direction normal to the shock. This behavior can be more easily observed by, for instance, plotting the limiter values along a horizontal line that crosses the upper shock of test case 3 at $z = 0.27$ m, as shown in Fig. 9.

## 5. CONCLUDING REMARKS

In this paper, the $R_3$ limiter function, recently introduced by Nishikawa (2022), is compared with the traditional Venkatakrishnan limiter (Venkatakrishnan, 1995) and one of its modified versions, as proposed by Wang (2000), in the context of a second-order, cell-centered FV scheme. The resulting numerical method is utilized to simulate a transonic flow around the two-dimensional NACA 0012 airfoil. Three distinct angles of attack are considered, producing three different shock wave patterns. Albeit different limiter formulations are considered, overall similar results are obtained in relation to the estimated physical property values. Nevertheless, by observing the values assumed by the pressure limiter in the regions that surround the airfoil surface, it is possible to conclude that, as expected, the $R_3$ limiter exhibits less dissipative characteristics than the other two formulations. This particular limiter is often activated only on the two cells that are directly adjacent to the shock wave. However, for the current application and numerical scheme order, such feature does not produce any overwhelming improvement in the quality of the solution. Hence, for second-order schemes that are similar in nature to the one used here, the usage of Wang's modified version of Venkatakrishnan's limiter, or even the traditional Venkatakrishnan limiter itself, is probably well suited for most applications. Wang's modified version, however, is theoretically more robust for situations in which very large differences in cell length scales are observed in the discretized domain, in contrast with the original formulation. That is not the case for the mesh employed here, though.

Furthermore, the authors observed that, provided the limiter control parameters are kept within recommended interval values, changes made to those parameters produce very little effect over the properties of the flow field, being well within engineering tolerance levels. Efficient residue convergence to machine zero does not appear to be possible in the current test cases, likely requiring the implementation of a tweaked nonlinear solver in order to improve its convergence properties. Finally, the total absence of a limiter formulation generates spurious oscillations in the region near the shock wave. These high-amplitude oscillations are advected throughout the domain, which excite unstable nonlinear numerical modes and make the solution to the discrete system of partial differential equations quickly diverge.

## 6. ACKNOWLEDGMENTS

The authors wish to express their gratitude to the São Paulo Research Foundation, FAPESP, which has supported the present research under the Research Grants No. 2021/00147-8 and No. 2013/07375-0. The authors also gratefully acknowledge the support for the present research provided by Conselho Nacional de Desenvolvimento Científico e Tecnológico, CNPq, under the Research Grant No. 309985/2013-7. The work is further supported by the computational resources of the Center for Mathematical Sciences Applied to Industry (CeMEAI), also funded by FAPESP under the

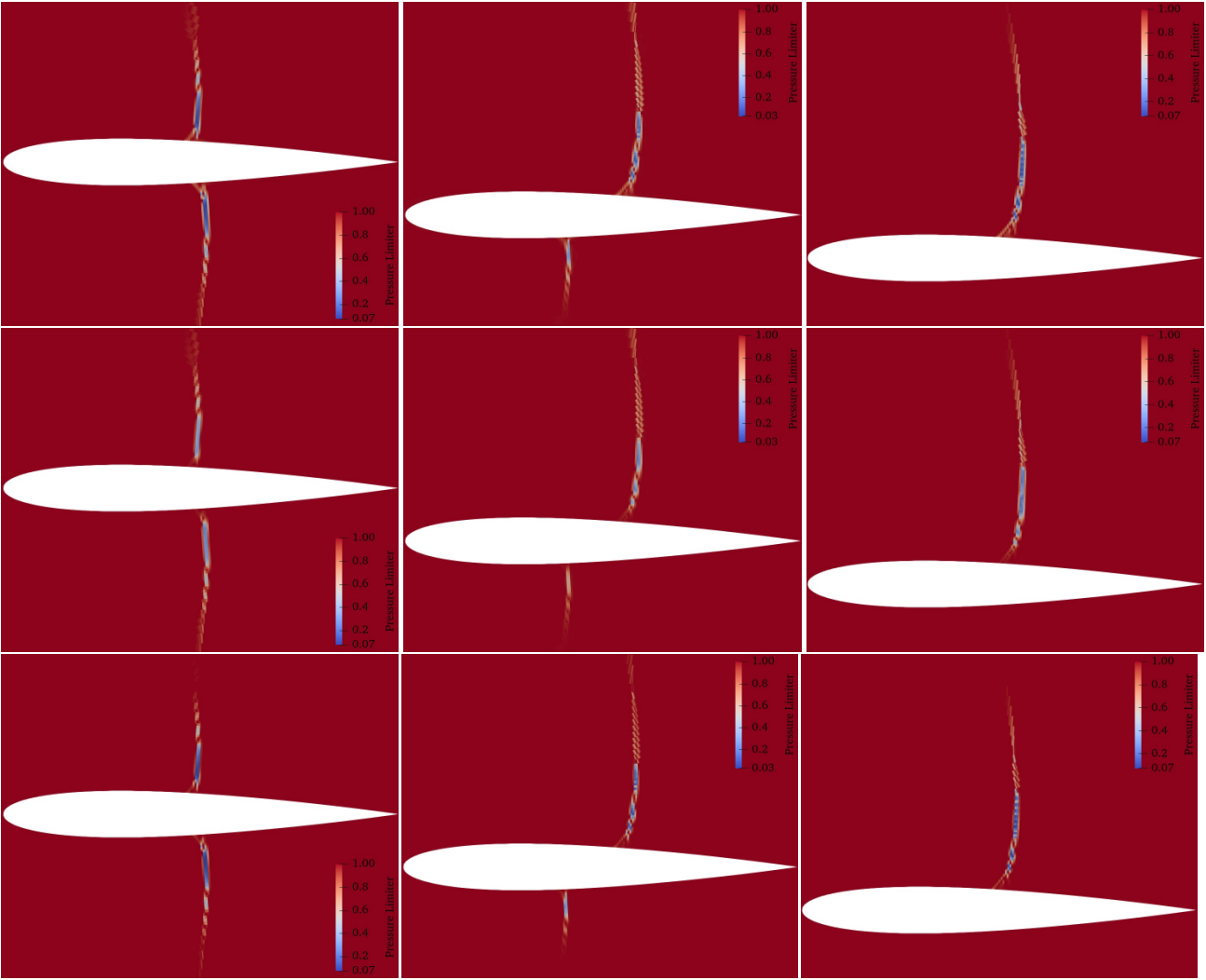

Figure 8. Pressure limiter contours for all combinations of test case and limiter formulation. Results from each row are obtained using the $\psi^V$, $\psi^W$ and $\psi^{R_3}$ limiters, respectively. Likewise, results from each column refer to test cases 1, 2 and 3, respectively.

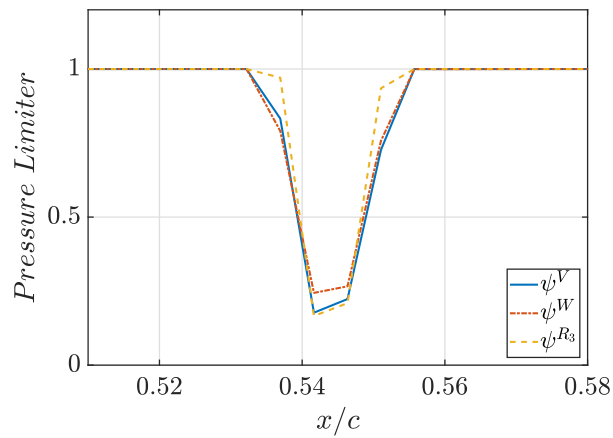

Figure 9. Pressure limiter values computed along a horizontal line, with $z = 0.27$ m, for test case 3. Focus is given to the region that surrounds the upper surface shock wave. Sample points are attributed to the cell centroids for plotting purposes.

Research Grant No. 2013/07375-0.

# 8. RESPONSIBILITY NOTICE

The authors are solely responsible for the printed material included in this paper.